\pdfoutput=1
\documentclass[a4paper,cits,notoc]{JINST}
\usepackage{microtype}

\newcommand{\bb}{\ensuremath{\beta\beta}}
\newcommand{\bbonu}{\ensuremath{\beta\beta0\nu}}
\newcommand{\bbtnu}{\ensuremath{\beta\beta2\nu}}
\newcommand{\mbb}{\ensuremath{m_{\beta\beta}}}

\newcommand{\ckky}{\ensuremath{\rm counts/(keV \cdot kg \cdot y)}}

\newcommand{\Qbb}{\ensuremath{Q_{\beta\beta}}}

\newcommand{\XE}{\ensuremath{{}^{136}\rm Xe}}
\newcommand{\GE}{\ensuremath{{}^{76}\rm Ge}}

\newcommand{\TL}{\ensuremath{{}^{208}\rm{Tl}}}

\newcommand{\BI}{\ensuremath{{}^{214}}Bi}

\title{NEXT, high-pressure xenon gas experiments for ultimate sensitivity to Majorana neutrinos}

\author{J.J.~G\'omez-Cadenas,\thanks{Corresponding author.}~ J.~Mart\'in-Albo and F.~Monrabal for the NEXT Collaboration \\
Instituto de F\'isica Corpuscular (IFIC), CSIC \& Universidad de Valencia\\
Calle Catedr\'atico Jos\'e Beltr\'an, 2, 46980 Paterna, Valencia, Spain\\
E-mail: \email{gomez@mail.cern.ch, justo.martin-albo@ific.uv.es, francesc.monrabal@ific.uv.es}}
  
\abstract{
In this paper we describe an innovative type of Time Projection Chamber (TPC), which uses high-pressure xenon gas (HPXe) and electroluminescence amplification of the ionization charge as the basis of an apparatus capable of fully reconstructing the energy and topological signature of rare events.
We will discuss a specific design of such HPXe TPC, the NEXT-100 detector, that will search for \bbonu\ events using 100--150 kg of xenon enriched in the isotope \XE. NEXT-100 is currently under construction, after completion of an accelerated and very successful R\&D period. It will be installed at the Laboratorio Subterr\'aneo de Canfranc (LSC), in Spain. The commissioning run is expected for late 2013 or early 2014.
We will also present physics arguments that suggest that the HPXe technology can be extrapolated to the next-to-next generation (e.g, a fiducial mass of 1 ton of target), which will fully explore the Majorana nature of the neutrino if the mass hierarchy is inverse. 
}


\begin{document}

%%%%%%%%%%%%%%%%%%%%%%%%%%%%%%%%%%%%%%%%%%%%%%%%%%%%%%%%%%%%
\section{Neutrinoless double beta decay and Majorana neutrinos}
Double beta decay (\bb) is a very rare nuclear transition in which a nucleus with $Z$ protons decays into a nucleus with $Z+2$ protons and the same mass number $A$. It can only be observed in those isotopes where the decay through the $\beta$ channel is forbidden or highly suppressed. There are 35 naturally-occurring isotopes that fulfill such a condition. Two decay modes are usually considered:
%%%
\begin{itemize}
\item The standard two-neutrino mode (\bbtnu), consisting in two simultaneous beta decays, $(Z,A) \rightarrow\ (Z+2,A) + 2\ e^{-} + 2\ \overline{\nu}_{e}$, which has been observed in several isotopes with typical half-lives in the range of $10^{18}$--$10^{21}$ years. 
\item The neutrinoless mode (\bbonu), $(Z,A) \rightarrow (Z+2,A) + 2\ e^{-}$, which violates lepton-number conservation, and is therefore forbidden in the Standard Model of particle physics. No convincing experimental evidence of this decay exists to date.
\end{itemize}
%%%

The implications of experimentally establishing the existence of \bbonu\ would be profound \cite{GomezCadenas:2011it}. First, it would demonstrate that total lepton number is violated in physical phenomena, an observation that could be linked to the cosmic asymmetry between matter and antimatter through a process known as \emph{leptogenesis} \cite{Fukugita:1986hr, Davidson:2008bu}. Second, the discovery of \bbonu\ would establish a Majorana nature for the neutrino \cite{Schechter:1981bd}. Majorana neutrinos provide a natural explanation to the smallness of neutrino masses, the so-called \emph{seesaw mechanism} \cite{Minkowski:1977sc, GellMann:1980vs, Yanagida:1979, Mohapatra:1979ia, Schechter:1980gr}.

Several underlying mechanisms --- involving, in general, physics beyond the Standard Model  --- have been proposed for \bbonu, the simplest one being the virtual exchange of light Majorana neutrinos. Assuming this to be the dominant one at low energies, the half-life of \bbonu\ can be written as
%%%
\begin{equation}
(T^{0\nu}_{1/2})^{-1} = G^{0\nu} \ \big|M^{0\nu}\big|^{2} \ \mbb^{2}.
\label{eq:Tonu}
\end{equation}
%%%
In this equation, $G^{0\nu}$ is an exactly-calculable phase-space integral for the emission of two electrons; $M^{0\nu}$ is the nuclear matrix element (NME) of the transition, that has to be evaluated theoretically using nuclear models; and \mbb\ is the \emph{effective Majorana mass} of the electron neutrino:
\begin{equation}
\mbb = \Big| \sum_{i} U^{2}_{ei} \ m_{i} \Big| \ ,
\end{equation}
where $m_{i}$ ($i=1,2,3$) are the neutrino mass eigenstates and $U_{ei}$ are elements of the neutrino mixing matrix. Therefore a measurement of the \bbonu\ decay rate would provide direct information on neutrino masses.

\begin{figure}
\centering
\includegraphics[width=0.60\textwidth]{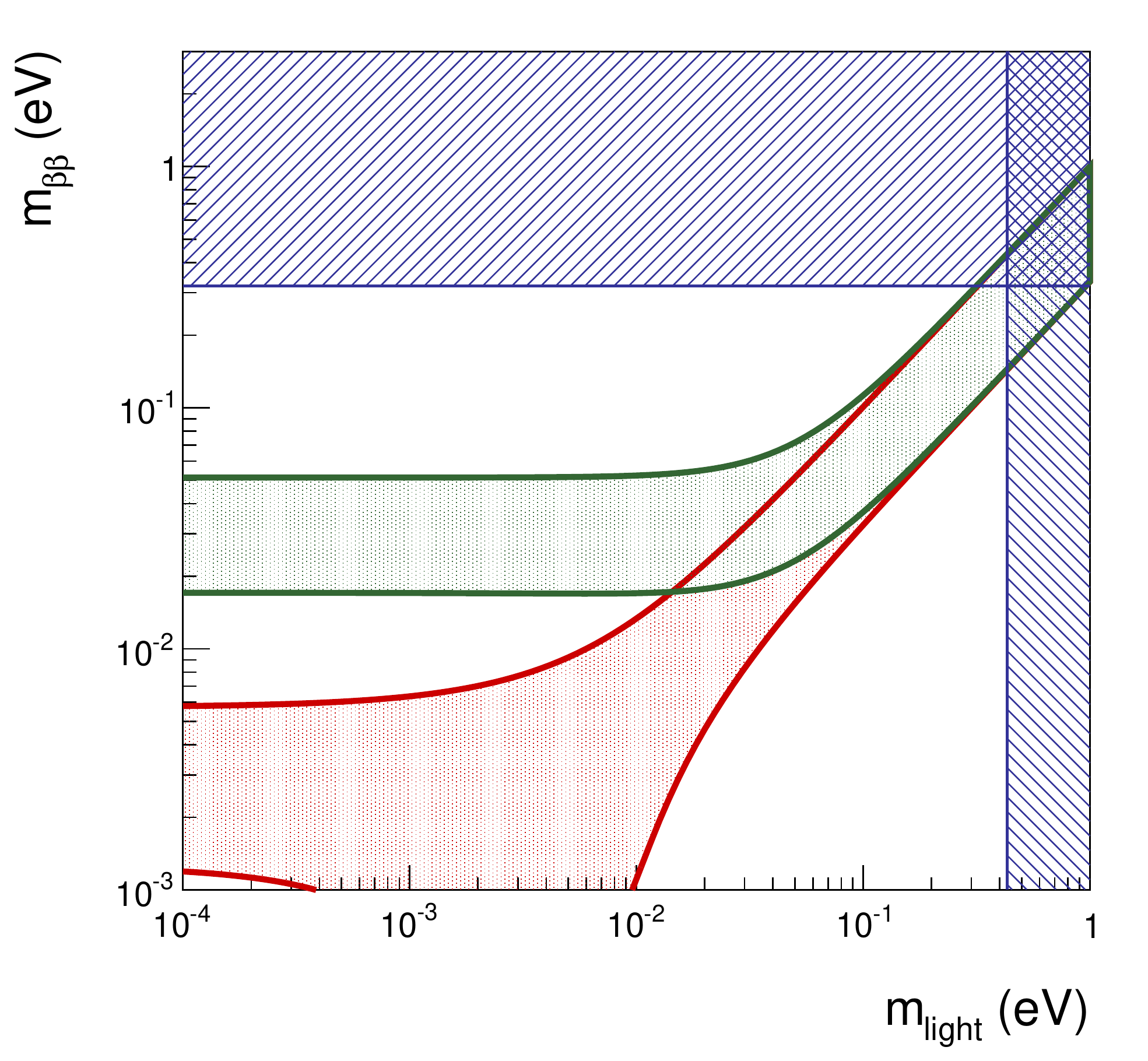}
\caption{The effective neutrino Majorana mass, \mbb, as a function of the lightest neutrino mass, $m_{\rm light}$. The green band corresponds to the inverse hierarchy of neutrino masses, whereas the red one corresponds to the normal ordering. The vertically-excluded region comes from cosmological bounds; the horizontally-excluded one from \bbonu\ constraints.} \label{fig:mbetabetavsmlight}
\end{figure}

The relationship between \mbb\ and the actual neutrino masses $m_i$ is actually affected by the uncertainties in the measured oscillation parameters, the unknown neutrino mass ordering (normal or inverted), and the unknown phases in the neutrino mixing matrix (both Dirac and Majorana). For example, the relationship between \mbb\ and the lightest neutrino mass $m_{\rm light}$ (which is equal to $m_1$ or $m_3$ in the normal and inverted mass orderings, respectively) is illustrated in figure \ref{fig:mbetabetavsmlight}. 

%%%%%
%%%%%

The horizontally-excluded region in the figure corresponds to the experimental constraint set by the Heidelberg-Moscow (HM) experiment, which was until very recently the most sensitive limit to the half-life of \bbonu\ so far: $T^{0\nu}_{1/2}(\GE) \ge 1.9\times10^{25}$ years at 90\% CL \cite{KlapdorKleingrothaus:2000sn}. Notice that a subgroup of the HM experiment interprets the data as {\em evidence} of a positive signal, with a best value for the half-life of $1.5\times10^{25}$ years \cite{KlapdorKleingrothaus:2001ke}, corresponding to an effective Majorana mass of about 400 meV . This claim  was very controversial, and still awaits a definitive experimental response.

%%%%%%%%%%%%%%%%%%%%%%%%%%%%%%%%%%%%%%%%%%%%%%%%%%%%%%%%%%%%
\section{Experimental aspects}
The discovery of \bbonu\ would mark a breakthrough in particle physics. A single, unequivocal observation of the decay would prove the Majorana nature of neutrinos and the violation of lepton number.  Alas, this is not, by any means, an easy task. The design of a detector capable of identifying efficiently and unambiguously such a rare signal poses a major experimental problem.

The detectors used in double beta decay searches are designed, in general, to measure the energy of the radiation emitted by a \bb\ source. In the case of \bbonu, the sum of the kinetic energies of the two released electrons is always the same, and is equal to the mass difference between the parent and the daughter nuclei: $Q_{\bb} \equiv M(Z,A)-M(Z+2,A)$. However, due to the finite energy resolution of any detector, \bbonu\ events are reconstructed within a non-zero energy range centered around \Qbb, typically following a gaussian distribution. Other processes occurring in the detector can fall in that region of energies, becoming a background and compromising drastically the experiment's expected sensitivity.

All double beta decay experiments have to deal with an intrinsic background, the \bbtnu, that can only be suppressed by means of good energy resolution. Backgrounds of cosmogenic origin force the underground operation of the detectors. Natural radioactivity emanating from the detector materials and surroundings can easily overwhelm the signal peak, and hence careful selection of radiopure materials is essential. Additional experimental signatures that allow the distinction of signal and background are a bonus to provide a robust result.

Besides energy resolution and control of backgrounds, several other factors such as detection efficiency or the scalability to large masses must be also taken into account during the design of a double beta decay experiment. The simultaneous optimization of all these parameters is most of the time conflicting, if not impossible, and consequently many different experimental approaches have been proposed. In order to compare them, a figure of merit, the experimental sensitivity to \mbb, is normally used \cite{GomezCadenas:2010gs}:
%%%
\begin{equation}
\mbb = K \ \sqrt{1/\varepsilon} \ \left(\frac{b\cdot\Delta E}{M\cdot t} \right)^{1/4}\, ,
\label{eq:mbb}
\end{equation}
%%%
where $\varepsilon$ is the detection efficiency, $\Delta E$ is the energy resolution window where the \bbonu\ signal will be reconstructed, $b$ is the background rate (in counts per year, kilogram of \bb\ isotope and keV) in the region of interest, $M$ is the \bb\ isotope mass, and $t$ is the data-taking time.

%%%%%%%%%%%%%%%%%%%%%%%%%%%%%%%%%%%%%%%%%%%%%%%%%%%%%%%%%%
\section{The current generation of double beta decay experiments}
The observation of neutrino oscillations \cite{Fukuda:1998mi, GonzalezGarcia:2007ib, Beringer:1900zz}, which demonstrated that neutrinos are massive particles (an essential condition for \bbonu\ to exist), and the HM result prompted a new generation of \bbonu\ experiments that promises to push the current limits down to neutrino masses of about 100 meV. The status of the field has recently been reviewed \cite{GomezCadenas:2011it}. Among the proposed and on-going experiments, one can find many different experimental techniques, each one with its pros and cons. Some emphasize the energy resolution and the detection efficiency, like the germanium calorimeters (GERDA \cite{Cattadori:2012fy} and {\sc Majorana} \cite{Wilkerson:2012ga}), which will also put the HM claim to test shortly. Bolometers such as CUORE \cite{Gorla:2012gd} also offer excellent energy resolution, and, in addition, the possibility of deploying large masses of \bb\ isotope using natural tellurium.  

The first new-generation experiments that have produced results are xenon-based: KamLAND-Zen \cite{KamLANDZen:2012aa}, in which xenon is dissolved in liquid scintillator; and EXO-200 \cite{Auger:2012gs}, a liquid xenon (LXe) TPC operating at WIPP (USA). 
Indeed, xenon is an interesting isotope for \bbonu\ searches. Two naturally-occurring isotopes of xenon can decay \bb, $^{134}$Xe ($\Qbb=825$~keV) and \XE\ ($\Qbb=2458$~keV). The latter, having a higher $Q$-value, is preferred because the decay rate is proportional to $\Qbb^{5}$, and the radioactive backgrounds are less abundant at higher energies. Besides, the \bbtnu\ mode of \XE\ is slow ($\sim2.3\times10^{21}$~years), and hence the experimental requirement for good energy resolution is less stringent than for other \bb\ sources. The process of isotopic enrichment is relatively simple and cheap compared to that of other \bb\ isotopes, and consequently \XE\ is the most obvious candidate for a future multi-ton \bb\ experiment.

EXO-200 has recently set the most stringent limit in the search for neutrinoless double beta decay \cite{Auger:2012ar}. It has searched for \bbonu\ events with a total exposure of 120.7 days and an active mass of 98.5 kg, which corresponds to 79.4 kg of \XE. The total exposure is 32.5 kg$\cdot$y. The background model predicts 4 events in the region of 1$\sigma$~around $\Qbb$ and 7.5 events in the 2$\sigma$ region. They observe 1 event in the 1$\sigma$ region of interest (ROI) and 5 events in the 2$\sigma$~ROI. A limit on the half-life of \bbonu\ is extracted from this observation: $T_{1/2}^{0\nu}(\XE) > 1.6 \times 10^{25}$ years. This limit is considerably better than the expected sensitivity of EXO-200 (e.g, the probability of observing one background event when 4 are expected is only 5\%). In terms of the effective neutrino mass, the EXO Collaboration quotes a sensitivity ranging between 140 and 380 meV, depending on the NME. 

EXO-200 deploys a total mass of 200 kg of enriched liquid xenon (85\% of \XE). About half of this mass is used for self-shielding. The energy resolution quoted by the Collaboration in their recent paper is 4\% FWHM at \Qbb. This is achieved by using the anti-correlation between the ionization and scintillation signals provided by the xenon. The background rate measured in the \emph{region of interest} (ROI) is $ 1.5 \times 10^{-3}\ckky$.

Notice that in order to fully cover the degenerate hierarchy one needs to improve the sensitivity of EXO-200 by about a factor of 10 in \mbb\ (from 200 to 20 meV), which, in turn, requires, using equation~(\ref{eq:mbb}), an increase of $10^4$ in mass if no other parameters are changed. This is clearly not an option, and therefore one needs to consider ways to improve the energy resolution and/or the background rate. This may prove difficult in LXe (the achievable resolution is limited by physics to a 3--3.5 \% FWHM at \Qbb, and the level of self-shielding by the fact that one needs to waste expensive enriched xenon), and, as we will discuss in the following sections, gaseous xenon may show better prospects.

%%%%%%%%%%%%%%%%%%%%%%%%%%%%%%%%%%%%%%%%%%%%%%%%%%%%%%%%%%%%
\section{The NEXT experiment and its innovative concepts}
The \emph{Neutrino Experiment with a Xenon TPC}\footnote{\href{http://next.ific.uv.es/}{http://next.ific.uv.es/}} (NEXT) will search for \bbonu\ in \XE\ using a high-pressure xenon gas (HPXe) time projection chamber (TPC). Such a detector 
offers major advantages for the search of neutrinoless double beta decay; namely: (a) \emph{good energy resolution}, better than $\sim1\%$ FWHM at \Qbb; (b) \emph{tracking capabilities} that provide a powerful signature to discriminate between signal (two electron tracks with a common vertex) and background (mostly, single electrons); (c) a \emph{fully active and homogeneus} detector; and (d) \emph{scalability} of the technique to larger masses of source isotope.

\begin{figure}
\centering
\includegraphics[width=0.65\textwidth]{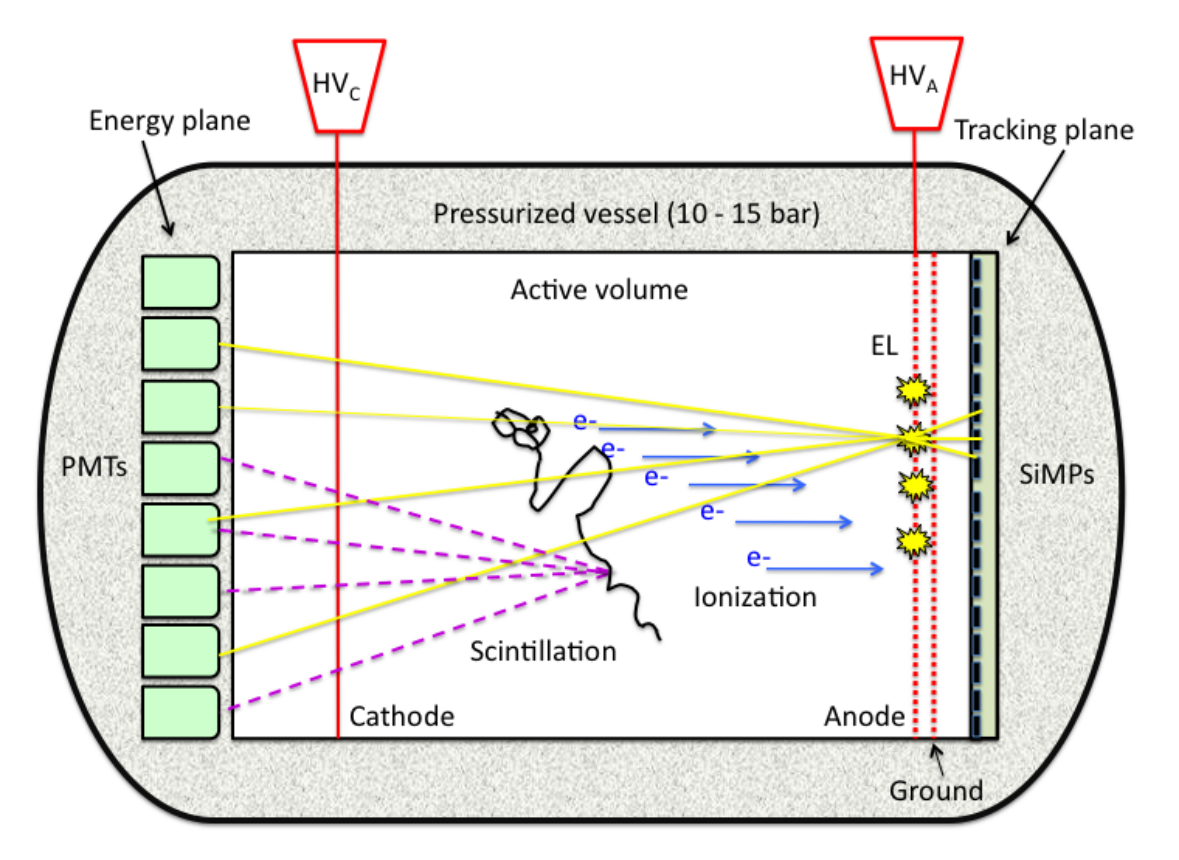}
\caption{The detection process in the NEXT TPC: EL light generated at the anode is recorded in the photosensor plane right behind it and used for tracking. It is also recorded in the photosensor plane behind the transparent cathode and used for a precise energy measurement. } \label{fig.SOFT}
\end{figure}

The design of NEXT is optimized for energy resolution by using proportional electroluminescent (EL) amplification of the ionization signal. The detection process, illustrated in figure~\ref{fig.SOFT}, is as follows. Particles interacting in the HPXe transfer their energy to the medium through ionization and excitation. The excitation energy is manifested in the prompt emission of VUV (178 nm) scintillation light. The ionization tracks (positive ions and free electrons) left behind by the particle are prevented from recombination by an electric field ($\sim0.5$ kV/cm at 10 bar). Negative charge carriers drift toward the TPC anode, entering a region, defined by two highly-transparent meshes, with an even more intense electric field ($\sim25$ kV/cm at 10 bar). There, further VUV photons are generated isotropically by electroluminescence. Therefore both scintillation and ionization produce an optical signal, to be detected with a sparse plane of PMTs (the \emph{energy plane}) located behind the cathode. The detection of the primary scintillation light constitutes the start-of-event, whereas the detection of EL light provides an energy measurement. Electroluminescent light provides tracking as well, since it is detected also a few mm away from production at the anode plane, via a dense array (1 cm pitch) of 1-mm$^{2}$ SiPMs (the \emph{tracking plane}).

%%%%%
%%%%%

\begin{figure}
\centering
\includegraphics[width=0.9\textwidth]{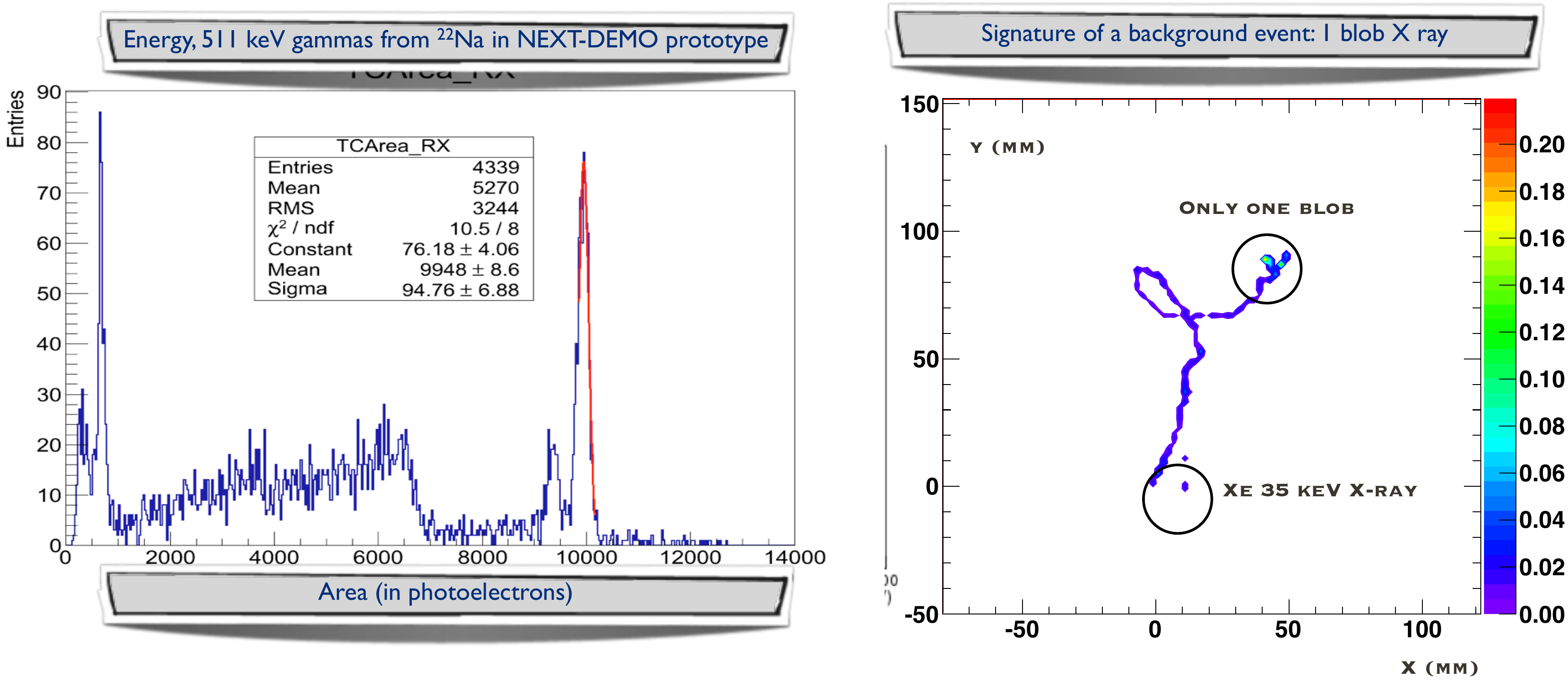} 
\caption{Left: NEXT has excellent energy resolution, as illustrated by the fit to the photo peak in the Na-22 spectrum (NEXT-DEMO data), which yields a resolution of 2 \% rms (this extrapolates to 1\% FWHM at \Qbb). Right: In addition NEXT has a topological signature, not available in most \bbonu\ detectors. The panel shows the reconstruction of a Monte Carlo background event. The background has only one electron (one blob) and the associated emission of a 35 keV X-ray. The color codes energy deposition in the TPC. } \label{fig.trk}
\end{figure}

Neutrinoless double beta decay events leave a distinctive topological signature in HPXe: a continuos track with larger energy depositions (\emph{blobs}) at both ends due to the Bragg-like peaks in the d$E$/d$x$ of the stopping electrons. In contrast, background electrons are produced by Compton or photoelectric interactions, and are characterized by a single blob and, often, by a satellite cluster corresponding to the emission of $\sim30$-keV fluorescence x-rays by xenon (see figure \ref{fig.trk}, right).

During the last three years, the NEXT Collaboration has developed an R\&D program with the specific goal of proving the performance of the technology, including energy resolution (see figure \ref{fig.trk}), tracking and stable operation. This program has resulted in the construction and operation of the NEXT-DEMO (installed at IFIC) and NEXT-DBDM (installed at the Lawrence Berkeley National Laboratory, USA) prototypes and is largely completed. Initial results of the NEXT-DEMO prototype have been presented at this conference \cite{Ferrario:2012}.

%%%%%
%%%%%

\begin{figure}
\centering
\includegraphics[width=0.9\textwidth]{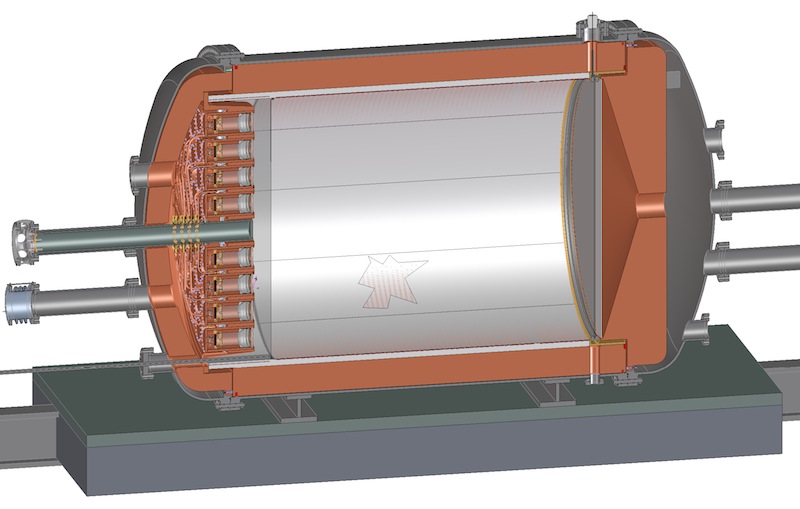}
\caption{Cross section of the NEXT-100 detector. The pressure vessel, 2.3 m long and 1.5 m diameter can hold up to 150 kg of xenon at 15 bar.}\label{fig:NEXT100}
\end{figure}

As a result of the R\&D phase, the NEXT Collaboration has published a \emph{Technical Design Report} \cite{Alvarez:2012haa} in 2012 describing a 100--150 kg experiment to be carried out, starting in 2014, at the Laboratorio Subterr\'aneo de Canfranc (LSC) in Spain. Figure \ref{fig:NEXT100} shows a drawing of the NEXT-100 detector.

%%%%%
%%%%%

NEXT-100 has the structure of a Matryoshka (a Russian nesting doll). The outermost layer of is a shield made of lead, which attenuates the background from the LSC rock by 6 orders of magnitude (e.g, the \TL\ photons are attenuated from $\sim 10^{12}$~per year to $\sim 10^{6}$~per year). The pressure vessel, built out of steel, is the next layer. Finally, an inner copper shield, 12 cm thick, constitutes the innermost and more radio-clean layer of the Matryoshka. In addition, all NEXT components have been selected and screened for low background. Of particular importance are the PMTs, whose activity is only 0.4 mBq of \BI\ and 0.3 mBq of \TL\ per unit. Our TDR shows a full quantification of the different contributions to the NEXT radioactive budget. 

The tracking capabilities of NEXT and its excellent energy resolution result in a very good background rejection factor, estimated to be $\sim 2\times 10^{-7}$. This translates into an excellent background rate of about $8 \times 10^{-4}$~\ckky.

\begin{figure}
\centering
\includegraphics[width=0.65\textwidth]{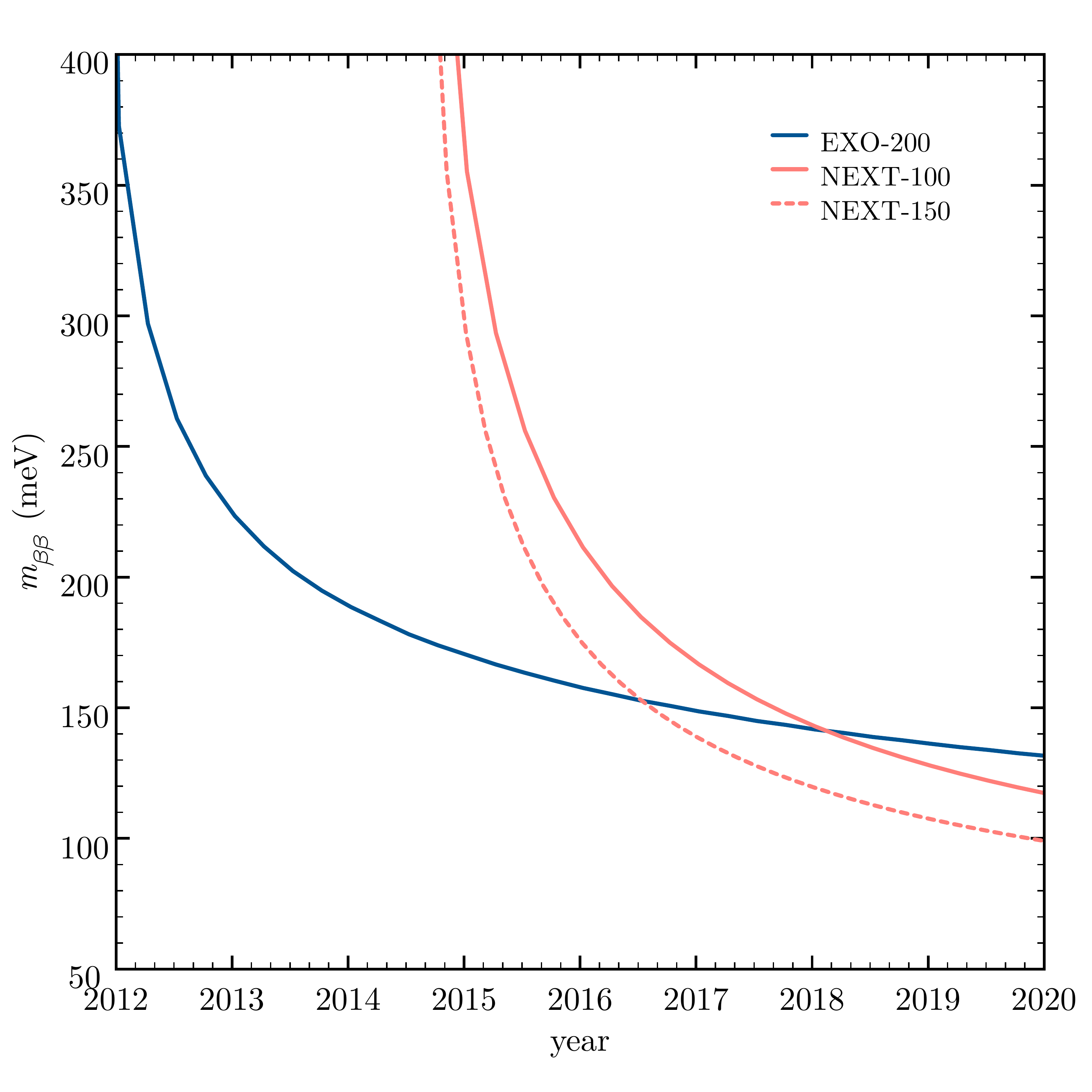}
\caption{Sensitivity of NEXT-100 versus the sensitivity of EXO-200. It is assumed that the NEXT-100 \emph{physics run} starts in mid 2014.} \label{fig.exoNext}
\end{figure}

The combination of excellent energy resolution, topological signature, large mass and a radio-clean detector results in an experiment with an excellent physics potential. Figure \ref{fig.exoNext} compares the sensitivity of NEXT-100 with that of EXO-200, assuming the parameters described in EXO-200 recent paper \cite{Auger:2012ar} and in NEXT-100 TDR \cite{Alvarez:2012haa}, and considering also the possibility that NEXT-100 runs with 100 or 150 kg of enriched xenon (the detector is designed to run at 10 bar with 100 kg of xenon, or at 15 bar with 150 kg). The calculation assumes a conservative overall exposure efficiency of 50\%, and that the NEXT-100 detector will start its physics run in mid 2014. A resolution of 4\% FWHM is used for EXO-200, and a conservative 1\% FWHM is assumed for NEXT-100. The sensitivity of NEXT-100 would cross that of EXO-200 after 2 years of running if 150 kg are deployed, or after some 3 years for 100 kg. 

This has two consequences. The first one is that, if a discovery is made by EXO-200 or any other new-generation experiment, NEXT-100 will be in an optimal position to check the claim, adding extra handles such as the topological signature. The second one is that, if no discovery is made by these experiments, NEXT has a window of opportunity for a discovery, in particular if 150 kg of mass are deployed. 

%%%%%
%%%%%

%%%%%%%%%%%%%%%%%%%%%%%%%%%%%%%%%%%%%%%%%%%%%%%%%%%%%%%%%%%%
\section{Towards a ton-scale version of NEXT}

As discussed above, if no discovery is made by the current generation of experiments, the full exploration of the inverse hierarchy of neutrino masses requires detectors of larger mass (at least 1 ton) and extremely low specific background ($\sim 10^{-4}$\ckky).
We argue that the potential of NEXT is extraordinary  for the next generation of \bbonu\ experiments. Specifically, we argue that NEXT can achieve the following:
%%%
\begin{itemize}
\item \emph{Energy resolution of 0.5\% FWHM at \Qbb}. As shown already by our prototypes in a restricted fiducial region. Operating NEXT-100 will provide the needed know-how to achieve such resolution in the whole active volume.
\item \emph{Low specific background} ($\sim 10^{-4}$~\ckky). The first step to achieve this very low background rate is to demonstrate that NEXT-100 can achieve its target of $8 \times 10^{-4}$~\ckky. At the same time, a redesign of the energy plane, moving the photomultipliers outside the fiducial volume and shielding them under a thick copper ring, would allow a drastic reduction of the background.
\item \emph{Scalability to the ton scale}. At 20 bar, the xenon can be fitted in 10 m$^3$. A symmetric TPC of 3 meters length and 1.1 meters radius would provide the required volume. This is a feasible extrapolation from NEXT-100, which is a symmetric TPC of 1.5 m length and 0.6 meters radius. Notice that, unlike almost any other \bb\ source, one ton of xenon can be acquired at a reasonable cost. In fact, one ton of enriched xenon already exists, combining the KamLAND-Zen, EXO and NEXT experiments. 
\end{itemize}
%%%

%%%%%
\begin{figure}
\centering
\includegraphics[width=0.65\textwidth]{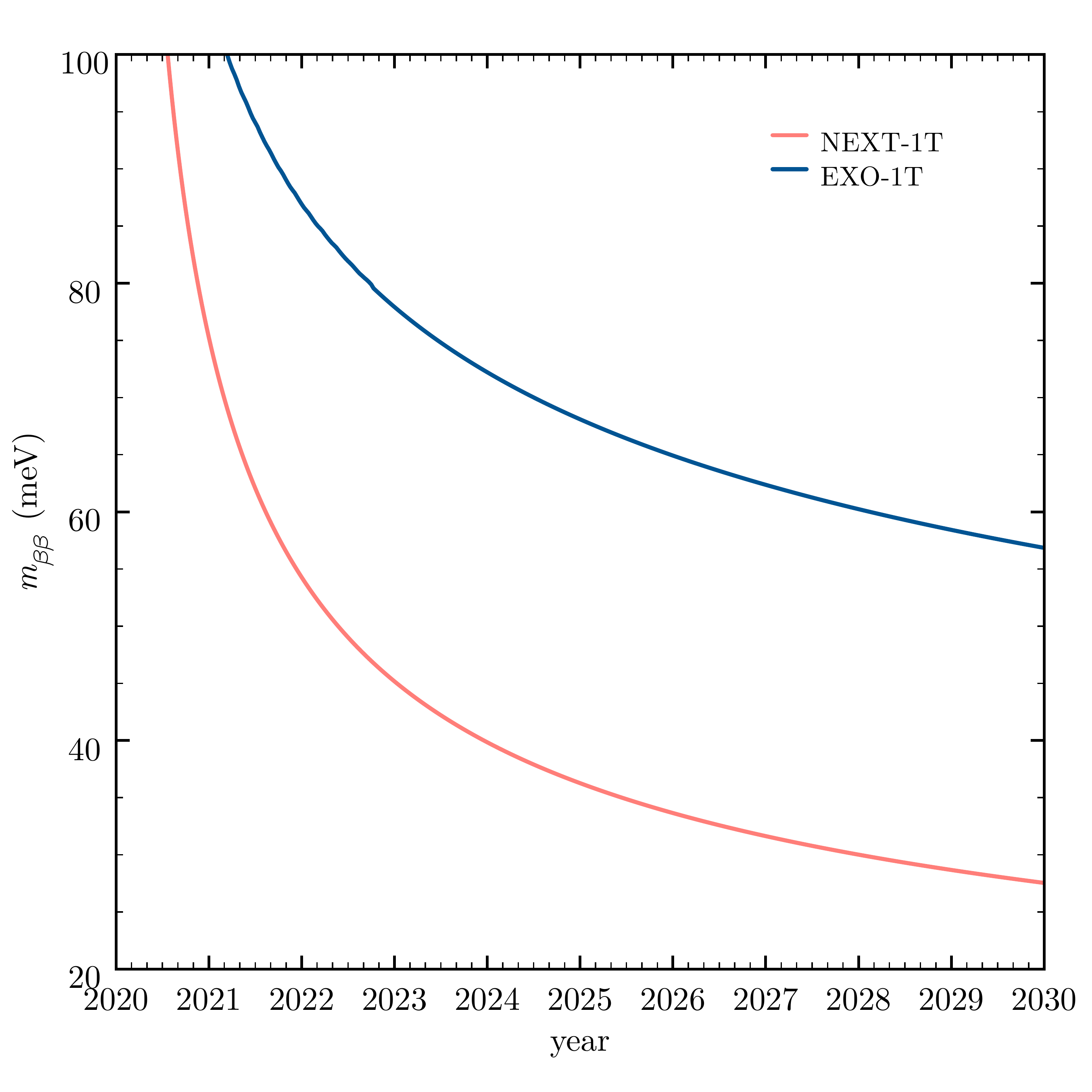}
\caption{The sensitivity of NEXT versus that of EXO in the 1-ton regime.} \label{fig.NextFuture}
\end{figure}
%%%%%

The advantages of HPXe over LXe can be argued in terms of equation (\ref{eq:mbb}). In an HPXe it is possible (as demonstrated by NEXT R\&D) to reach an ultimate energy resolution of 0.5\% FWHM at \Qbb\ and an ultimate background rate (thanks to the additional topological signature) of $\sim10^{-4}$. Thus an improvement of roughly two orders of magnitude with respect to EXO can be achieved. The remaining factor of $10^2$~needed to fully cover the inverse hierarchy needs to come from the exposure. This implies to increase the mass in at least a factor of 10. Improvements in detector efficiency and lifetime are needed to keep the running time to about 5 years. 

Figure \ref{fig.NextFuture} compares the sensitivity of NEXT with that of EXO, assuming improved parameters for both detectors in the 1-ton regime.  EXO resolution is still  4\%, while the resolution of NEXT is assumed to be improved to  0.5\% FWHM (as achieved by our prototypes). The background in the ROI for EXO is assumed to be  $ 0.7 \times 10^{-3}\ckky$ (EXO best predictions) and the background in the ROI for NEXT is
$ 0.2 \times 10^{-3}\ckky$, as discussed above. Notice that, unlike the current design of EXO, the improved NEXT could cover the full inverse hierarchy if the operation parameters are achieved.

%%%%%%%%%%%%%%%%%%%%%%%%%%%%%%%%%%%%%%%%%%%%%%%%%%%%%%%%%%%%
\section{Summary and Outlook}
Neutrinoless double beta decay experiments are the only practical way to establish whether the neutrino is its own antiparticle. The current generation of experiments, with a sensitivity to effective neutrino masses around 100 meV, is already operating, with initial results from KamLAND-Zen and EXO-200. The NEXT-100 detector could contribute in a decisive way to this exploration, in spite of a relatively late start.

At the same time, exploring the inverse hierarchy will require large masses (circa 1 ton) and a background level in the regime of very few counts per ton-year. Many of the current technologies may be eliminated either by lack of resolution or by the difficulties to extrapolate to large masses. In contrast, the NEXT concept which combines very good resolution, a topological signature and a clean detector, could lead the way in the near future.

%%%%%%%%%%%%%%%%%%%%%%%%%%%%%%%%%%%%%%%%%%%%%%%%%%%%%%%%%%%%
\acknowledgments
This work was supported by the Ministerio de Econom\'ia y Competitividad of Spain under grants CONSOLIDER-Ingenio 2010 CSD2008-0037 (CUP) and FPA2009-13697-C04-04, and by the Portuguese FCT and FEDER through program COMPETE, project PTDC/FIS/103860/2008.

%%%%%%%%%%%%%%%%%%%%%%%%%%%%%%%%%%%%%%%%%%%%%%%%%%%%%%%%%%%%
\bibliographystyle{JHEP}
\bibliography{biblio}

\providecommand{\href}[2]{#2}\begingroup\raggedright\begin{thebibliography}{10}

\bibitem{GomezCadenas:2011it}
J.~J. G\'omez-Cadenas, J.~Martin-Albo, M.~Mezzetto, F.~Monrabal, and M.~Sorel,
  {\it {The search for neutrinoless double beta decay}},  {\em Riv.\ Nuovo
  Cim.} {\bf 35} (2012) 29--98, [\href{http://xxx.lanl.gov/abs/1109.5515}{{\tt
  arXiv:1109.5515}}].

\bibitem{Fukugita:1986hr}
M.~Fukugita and T.~Yanagida, {\it {Baryogenesis Without Grand Unification}},
  {\em Phys. Lett.} {\bf B174} (1986) 45.

\bibitem{Davidson:2008bu}
S.~Davidson, E.~Nardi, and Y.~Nir, {\it {Leptogenesis}},  {\em Phys. Rept.}
  {\bf 466} (2008) 105--177, [\href{http://xxx.lanl.gov/abs/0802.2962}{{\tt
  arXiv:0802.2962}}].

\bibitem{Schechter:1981bd}
J.~Schechter and J.~Valle, {\it {Neutrinoless Double beta Decay in
  SU(2)$\times$U(1) Theories}},  {\em Phys. Rev.} {\bf D25} (1982) 2951.

\bibitem{Minkowski:1977sc}
P.~Minkowski, {\it $\mu \rightarrow e \gamma$ at a rate of one out of 1-billion
  muon decays?},  {\em Phys. Lett.} {\bf B67} (1977) 421.

\bibitem{GellMann:1980vs}
M.~Gell-Mann, P.~Ramond, and R.~Slansky, {\it {Complex spinors and unified
  theories}},  in {\em Proceedings of the {Supergravity} Stony Brook Workshop
  (New York, USA)}, edited by P.~van Nieuwenhuizen and D.~Z.~Freedman, North
  Holland Publ. Co., 1979.

\bibitem{Yanagida:1979}
T.~Yanagida, {\it Horizontal symmetry and masses of neutrinos},  in {\em
  {Proceedings of the Workshop on the Baryon Number of the Universe and Unified
  Theories (Tsukuba, Japan)}}, edited by O.~Sawada and A.~Sugamoto, KEK Report
  No. 79-18, 1979.

\bibitem{Mohapatra:1979ia}
R.~N. Mohapatra and G.~Senjanovic, {\it {Neutrino Mass and Spontaneous Parity
  Violation}},  {\em Phys. Rev. Lett.} {\bf 44} (1980) 912.

\bibitem{Schechter:1980gr}
J.~Schechter and J.~Valle, {\it {Neutrino Masses in SU(2)$\times$U(1)
  Theories}},  {\em Phys.\ Rev.} {\bf D22} (1980) 2227.

\bibitem{KlapdorKleingrothaus:2000sn}
H.~Klapdor-Kleingrothaus, A.~Dietz, L.~Baudis, G.~Heusser, I.~Krivosheina,
  et~al., {\it {Latest results from the Heidelberg-Moscow double beta decay
  experiment}},  {\em Eur. Phys. J.} {\bf A12} (2001) 147--154,
  [\href{http://xxx.lanl.gov/abs/hep-ph/0103062}{{\tt hep-ph/0103062}}].

\bibitem{KlapdorKleingrothaus:2001ke}
H.~Klapdor-Kleingrothaus, A.~Dietz, H.~Harney, and I.~Krivosheina, {\it
  {Evidence for neutrinoless double beta decay}},  {\em Mod. Phys. Lett.} {\bf
  A16} (2001) 2409--2420, [\href{http://xxx.lanl.gov/abs/hep-ph/0201231}{{\tt
  hep-ph/0201231}}].

\bibitem{GomezCadenas:2010gs}
J.~J. Gomez-Cadenas, J.~Martin-Albo, M.~Sorel, P.~Ferrario, F.~Monrabal,
  et~al., {\it {Sense and sensitivity of double beta decay experiments}},  {\em
  JCAP} {\bf 1106} (2011) 007, [\href{http://xxx.lanl.gov/abs/1010.5112}{{\tt
  arXiv:1010.5112}}].

\bibitem{Fukuda:1998mi}
{\bf Super-Kamiokande Collaboration} Collaboration, Y.~Fukuda et~al., {\it
  {Evidence for oscillation of atmospheric neutrinos}},  {\em Phys.Rev.Lett.}
  {\bf 81} (1998) 1562--1567,
  [\href{http://xxx.lanl.gov/abs/hep-ex/9807003}{{\tt hep-ex/9807003}}].

\bibitem{GonzalezGarcia:2007ib}
M.~Gonzalez-Garcia and M.~Maltoni, {\it {Phenomenology with Massive
  Neutrinos}},  {\em Phys.\ Rept.} {\bf 460} (2008) 1--129,
  [\href{http://xxx.lanl.gov/abs/0704.1800}{{\tt arXiv:0704.1800}}].

\bibitem{Beringer:1900zz}
{\bf Particle Data Group} Collaboration, J.~Beringer et~al., {\it {Review of
  Particle Physics (RPP)}},  {\em Phys.\ Rev.} {\bf D86} (2012) 010001.

\bibitem{Cattadori:2012fy}
C.~M. Cattadori, {\it {GERDA status report: Results from commissioning}},  {\em
  J.Phys.Conf.Ser.} {\bf 375} (2012) 042008.

\bibitem{Wilkerson:2012ga}
J.~Wilkerson, E.~Aguayo, F.~Avignone, H.~Back, A.~Barabash, et~al., {\it {The
  MAJORANA demonstrator: A search for neutrinoless double-beta decay of
  germanium-76}},  {\em J.Phys.Conf.Ser.} {\bf 375} (2012) 042010.

\bibitem{Gorla:2012gd}
{\bf CUORE} Collaboration, P.~Gorla, {\it {The CUORE experiment: Status and
  prospects}},  {\em J.Phys.Conf.Ser.} {\bf 375} (2012) 042013.

\bibitem{KamLANDZen:2012aa}
{\bf KamLAND-Zen} Collaboration, {\it {Measurement of the double-$\beta$ decay
  half-life of $^{136}$Xe with the KamLAND-Zen experiment}},  {\em Phys.Rev.}
  {\bf C85} (2012) 045504, [\href{http://xxx.lanl.gov/abs/1201.4664}{{\tt
  arXiv:1201.4664}}].

\bibitem{Auger:2012gs}
M.~Auger, D.~Auty, P.~Barbeau, L.~Bartoszek, E.~Baussan, et~al., {\it {The
  EXO-200 detector, part I: Detector design and construction}},  {\em JINST}
  {\bf 7} (2012) P05010, [\href{http://xxx.lanl.gov/abs/1202.2192}{{\tt
  arXiv:1202.2192}}].

\bibitem{Auger:2012ar}
{\bf EXO} Collaboration, M.~Auger et~al., {\it {Search for Neutrinoless
  Double-Beta Decay in $^{136}$Xe with EXO-200}},  {\em Phys.Rev.Lett.} {\bf
  109} (2012) 032505, [\href{http://xxx.lanl.gov/abs/1205.5608}{{\tt
  arXiv:1205.5608}}].

\bibitem{Ferrario:2012}
{\bf NEXT} Collaboration, P.~Ferrario, {\it Results of the next-demo
  prototype},  {\em JINST} (2012). Proceedings of iWoRID2012.

\bibitem{Alvarez:2012haa}
{\bf NEXT} Collaboration, V.~Alvarez et~al., {\it {NEXT-100 Technical Design
  Report (TDR): Executive Summary}},  {\em JINST} {\bf 7} (2012) T06001,
  [\href{http://xxx.lanl.gov/abs/1202.0721}{{\tt arXiv:1202.0721}}].

\end{thebibliography}\endgroup
\end{document}